# Signatures of chaos in the dynamics of quantum discord

Vaibhav Madhok,[1] Vibhu Gupta,[2,3] Angele M. Hamel,[1] and Shohini Ghose[1,3]

[1] *Department of Physics and Computer Science, Wilfrid Laurier University, Waterloo, Canada*[*]

[2] *Department of Physics and Astronomy, University of Waterloo, Canada*

[3] *Institute for Quantum Computing, University of Waterloo, Canada*[†]

We identify signatures of chaos in the dynamics of discord in a multiqubit system collectively modelled as a quantum kicked top. The evolution of discord between any two qubits is quasiperiodic in regular regions, while in chaotic regions, the quasiperiodicity is lost. As the initial wave function is varied from the regular regions to the chaotic sea, a contour plot of the time averaged discord remarkably reproduces the structures of the classical stroboscopic map. We also find surprisingly opposite behaviour of two-qubit discord versus two-qubit entanglement. Our calculations provide the first evidence of signatures of chaos in dynamically generated discord.

## INTRODUCTION

Classical chaos has been widely studied in a variety of contexts including weather patterns, population dynamics and chemical reactions [1]. Chaos in classical systems can be defined as hypersensitivity to initial conditions. Even very simple non-linear classical systems like the logistic map can exhibit chaotic dynamics. A natural question that arises is how to characterize chaos at the quantum level. This has led to the development of the field of quantum chaos - the study of how chaos manifests itself in the quantum regime. Signatures of chaos in quantum systems have been explored in the context of level statistics of chaotic Hamiltonians [2, 3], the dynamics of open quantum systems undergoing measurement or decoherence [4, 5] and hypersensitivity of a system to perturbations [6, 7]. More recently, there has been significant interest in the effect of chaos on quantum entanglement, both for a fundamental understanding of the quantum-classical connection and also due to the importance of entanglement as a resource for quantum information processing [8–11]. However, entanglement does not completely capture the quantum correlations of a system, and neither does it seem to be the only reason behind the quantum advantage in quantum information processing. Quantum discord aims to fill this gap and capture essentially all the quantum correlations in a quantum state using information theoretic measures [12, 13]. There is a considerable interest in quantum discord as recent studies show that it may account for the speedup in the performance of certain quantum algorithms compared to classical ones [14].

In this work, we provide the first evidence of signatures of chaos in the the dynamical behavior of discord in a quantum kicked top. An advantage of the quantum kicked top is that for a given angular momentum $j$, it can be regarded as a quantum simulation of a collection of $N = 2j$ spin-half particles whose evolution is restricted to the symmetric subspace under particle exchange. Thus, we have a multiqubit system where the collective behaviour of the qubits is governed by the the kicked top Hamiltonian. Another advantage of this approach is that it allows us to study discord between any two qubits and compare it to pairwise entanglement between two qubits or to bipartite entanglement between two qubits and the remaining qubits. Here, we present calculations showing various signatures of chaos in the dynamics of discord between any two qubits. The discord dynamics exhibits regular, quasiperiodic behaviour in a regular regime, but not in a chaotic regime. A contour plot of the time-averaged discord reproduces the classical phase space structures even in a deeply quantum regime. We find a surprising relationship between two-qubit discord and two-qubit concurrence - a measure of pairwise entanglement. When discord increases, concurrence decreases and vice versa. The two-qubit discord is robust and remains nonzero in a chaotic regime whereas the concurrence quickly decreases to zero. The quantum kicked top was experimentally realized recently [15], and our calculations are performed using parameters that are experimentally accessible using current technology.

## BACKGROUND

### Quantum kicked top

The quantum kicked top is described by the Hamiltonian [2, 15, 17]

$$H = \frac{\kappa}{2j\tau}{J_z}^2 + pJ_y\sum\delta(t - n\tau). \quad (1)$$

Here $J_x, J_y$ and $J_z$ are components of the angular momentum operator $\mathbf{J}$. The time between periodic kicks is given by $\tau$. Each kick is a rotation about the $y$ axis by an angle $p$. $\kappa$ is the strength of a twist applied between kicks, and is also the chaoticity parameter - as $\kappa$ is increased, the degree of chaoticity increases. Since the kick is in the form of delta kicks, we can express the Floquet map (evolution from kick to kick) as a sequence of operations given by,

$$U_\tau = \exp(-i\frac{\kappa}{2j\tau}J_z^2)\exp(-ipJ_y). \quad (2)$$

For a given value of angular momentum $j$, the Hilbert space dimension is $2j+1$. The finite dimension of the Hilbert space makes is possible to explore the dynamics without the need for truncation of the space.

The classical limit of this map can be obtained by writing the Heisenberg equations of motion for the expectation values of $J_x$, $J_y$ and $J_z$ and factorizing higher moments of the angular momentum operators in the limit of large $j$. The resulting equations describe the motion of an angular momentum vector on the surface of a sphere. The dynamics can be understood as a rotation by a fixed angle $p$ about the $y$-axis by angle $p$, followed by a rotation about the $z$ axis by an angle proportional to the z component of the angular momentum. This sequence of transformations can result in chaotic dynamics due to the lack of enough constants of motion. In our analysis, we fix $p = \pi/2$ and choose $\kappa$ to be our chaoticity parameter. As we vary $\kappa$ from 0 to 6, the classical limit of the dynamics change from highly regular to completely chaotic. In the quantum description, as the dynamics becomes globally chaotic, and for $j \gg 1$, the Hamiltonian can be modelled as a random matrix selected from the appropriate ensemble [2]. It is this randomness that leads to the analog of ergodic mixing for quantum systems.

We can think of the total angular momentum $j$ as the sum of the angular momenta of $N = 2j$ individual spin-half particles or qubits. The qubits are identical and the system remains unchanged under the exchange of any two qubits. Hence the state vector is restricted to a symmetric subspace spanned by the basis states $\{|j,m\rangle; (m = -j, -j+1, ..., j)\}$ with $j = N/2$.

In order to explore the quantum dynamics and compare to the classical limit, we must pick an initial condition for the dynamics. In the classical case, the initial condition is a set of coordinates $\theta$ and $\phi$ which specify the initial direction of the classical angular momentum vector. The uncertainty principle does not allow us to pick a corresponding quantum initial condition. Instead we construct a minimum uncertainty state vector such that the expectation values of $J_x$, $J_y$ and $J_z$ define a vector pointing along the direction $\theta, \phi$. Such states are the spin coherent states, which can be expressed as [17–20]

$$|\theta,\phi\rangle = R(\theta,\phi)|j,j\rangle; -\pi \le \phi \le \pi, 0 \le \theta \le \pi \quad (3)$$

where,

$$R(\theta,\phi) = \exp\{i\theta[J_x\sin\phi - J_y\cos\phi]\} \quad (4)$$

with the expectation value of $\mathbf{J}$ given by

$$\langle\theta,\phi|\mathbf{J}/j|\theta,\phi\rangle = (\sin\theta\cos\phi, \sin\theta\sin\phi, \cos\theta). \quad (5)$$

And the relative variance of $\mathbf{J}$ in a state $|\theta,\phi\rangle$ is [17],

$$(1/j^2)\{\langle\theta,\phi|\mathbf{J}^2|\theta,\phi\rangle - \langle\theta,\phi|\mathbf{J}|\theta,\phi\rangle^2 = 1/j \quad (6)$$

This is the minimum uncertainty possible from the angular momentum commutation relations, and approaches zero as $j$ becomes very large.

## MEASURES OF QUANTUM CORRELATIONS

### Entanglement

For a particular value of $j$, the system can be decomposed into $N = 2j$ qubits. To quantify correlations among these qubits, we trace out two qubits from the rest of the system [21]. The two qubit state thus obtained is a mixed state and the Von Neumann entropy of this two qubit state, defined as $E_V = -\mathrm{Tr}\,(\rho\ln\rho)$, captures how the two qubits are entangled with the rest of the qubits [22]. Pairwise entanglement between the two qubits, can be quantified by concurrence [23]. For the two qubit case, it has a simplified expression defined as

$$C = \max(0, \Lambda) \quad (7)$$

where $\Lambda = \lambda_1 - \lambda_2 - \lambda_3 - \lambda_4$, and $\lambda_i$ are the eigenvalues in decreasing order of the matrix $\rho(\sigma_2 \otimes \sigma_2)\rho^*(\sigma_2 \otimes \sigma_2)$. $\sigma_2$ is a Pauli matrix and $\rho^*$ is the complex conjugate of $\rho$.

### Quantum Discord

Quantum discord is a measure that extracts all quantum correlations, including entanglement in a quantum state [12]. The approach to do this is to remove the classical correlations from the total correlations in a system. A measure of total correlations in a bipartite quantum system $\rho_{AB}$, is the quantum mutual information,

$$I(A:B) = H(A) + H(B) - H(A,B), \quad (8)$$

where $H(\cdot)$ is the von Neumann entropy, $H(\rho) \equiv -\mathrm{Tr}\,(\rho\log\rho)$. A definition of mutual information for classical probability distributions based on Bayes' rule is

$$I(A:B) = H(A) - H(A|B). \quad (9)$$

Here, the conditional entropy $H(A|B)$ is the average of the Shannon entropies of $A$, conditioned on the values of $B$, and reflects the ignorance in $A$ given the state of $B$. In the quantum case, we can describe measurements on $B$ by a POVM set $\{\Pi_i\}$, such that the conditioned state of $A$ given outcome $i$ is

$$\rho_{A|i} = \mathrm{Tr}_B(\Pi_i\rho_{AB})/p_i, \quad p_i = \mathrm{Tr}_{A,B}(\Pi_i\rho_{AB}). \quad (10)$$

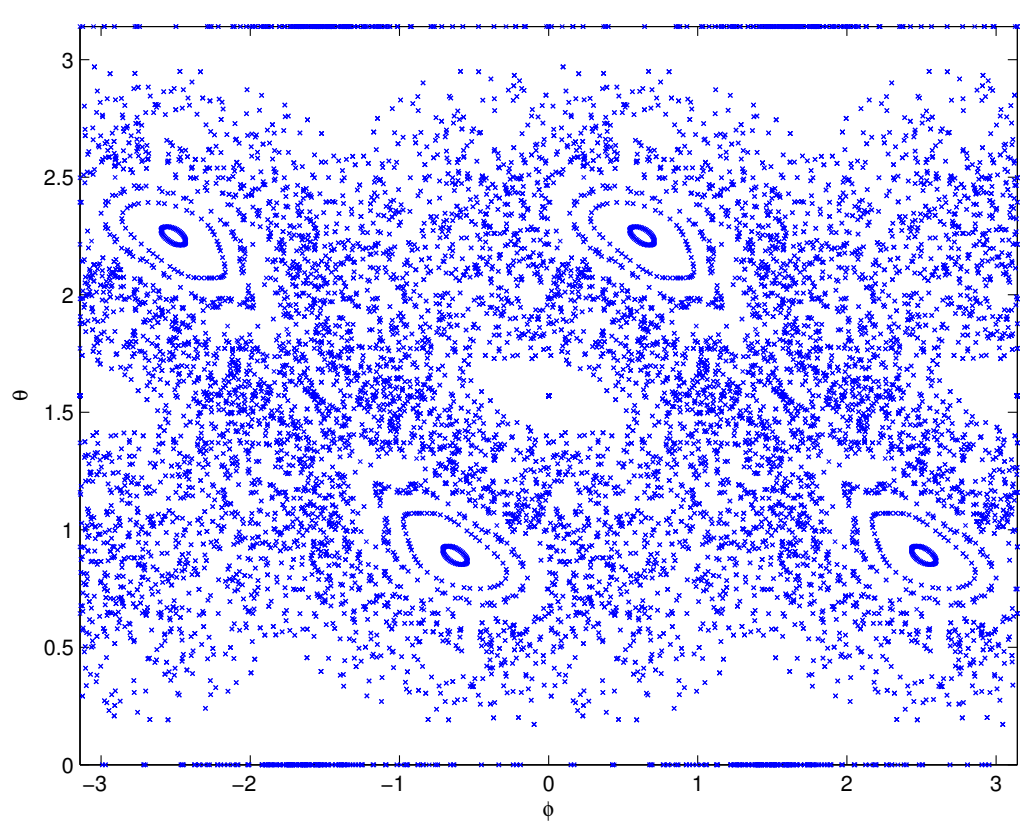

FIG. 1. Classical stroboscopic map for the kicked top. The direction of the angular momentum vector is plotted after each kick for different initial conditions with $p = \pi/2, \kappa = 3$.

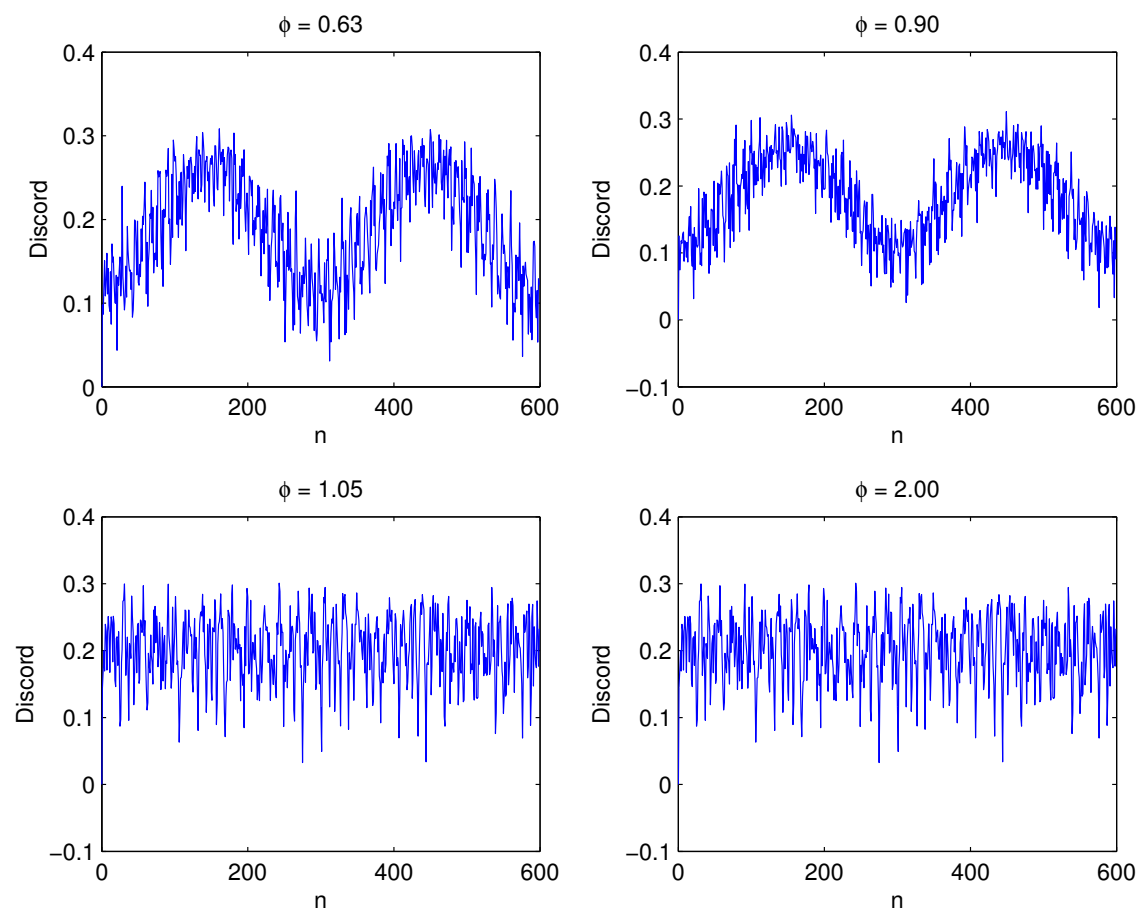

FIG. 2. Evolution of the two-qubit discord as a function of number of kicks for an initial spin coherent state centered on (a) Fixed point ($\theta = 2.25, \phi = 0.63$)(b) Regular island ($\theta = 2.25, \phi = 0.90$) (c) 'Edge of chaos' ($\theta = 2.25, \phi = 1.05$) (d) chaotic sea ($\theta = 2.25, \phi = 2.00$)

The corresponding entropy is then $\tilde{H}_{\{\Pi_i\}}(A|B) \equiv \sum_i p_i H(\rho_{A|i})$, from which one can write the quantum mutual information as $\mathcal{J}_{\{\Pi_i\}}(A : B) = H(A) - \tilde{H}_{\{\Pi_i\}}(A|B)$. Maximizing this over all $\{\Pi_i\}$, we obtain $\mathcal{J}(A : B) = \max_{\{\Pi_i\}}(H(A) - \tilde{H}_{\{\Pi_i\}}(A|B)) \equiv H(A) - \tilde{H}(A|B)$, where $\tilde{H}(A|B) = \min_{\{\Pi_i\}} \tilde{H}_{\{\Pi_i\}}(A|B)$. The minimum is achieved using rank 1 POVMs since the conditional entropy is concave over the set of convex POVMs [25]. Hence we arrive at a definition for quantum discord:

$$\begin{aligned} \mathcal{D}(A : B) &= I(A : B) - \mathcal{J}(A : B) \\ &= H(A) - H(A, B) + \min_{\{\Pi_i\}} \tilde{H}_{\{\Pi_i\}}(A|B), \end{aligned} \tag{11}$$

with $\{\Pi_i\}$ being rank 1 POVMs. Quantum discord is non-negative for all quantum states [12, 25], and it is subadditive [24].

## DYNAMICS OF QUANTUM CORRELATIONS

In order to study the connection between discord and chaos in the kicked top, we use the multiqubit representation of the system as discussed above. We trace out two qubits whose state we calculate after every application of the Floquet map [21] and calculate its discord. Since all the qubits are identical, this represents the discord between any two qubits in the system. We choose as initial states the minimum uncertainty spin coherent states, which can be characterized by angle $\theta$ and $\phi$. We take for the initial conditions, different points in the classical phase space for $p = \pi/2$ and $\kappa = 3$. (Fig. 1). We choose to take points corresponding to $(\theta, \phi)$=(2.25, 0.63), (2.25, 0.90), (2.25,1.05) and (2.25, 2.00). As can be seen from the classical phase space, these four points correspond to a fixed point, a point chosen in the regular island, 'edge of chaos' point, or a point on the boundary of the regular island and the chaotic sea, and a point chosen in the middle of the chaotic sea. The dynamics of discord for these specific four coordinates are shown in the Fig 2. For a coherent state initialized at the fixed point and in a regular island, discord increases at a slow rate and exhibits quasiperiodic oscillatory behaviour. For a spin coherent state initially in the chaotic sea ($\theta = 2.25, \phi = 0.63$), the discord increases more rapidly and reaches a quasi-steady state. The periodic modulation of discord dynamics is lost as the initial conditions are scanned from the regular region to the chaotic sea in the classical phase space. This indicates that there is a correlation between discord dynamics and regular versus chaotic regions of the classical phase space where the quantum state is initialized. It is remarkable that we see these signatures even though we are in a deeply quantum regime (j=4 which corresponds to just 8 qubits). Such a quantum regime is achievable in current experiments [15]. As we increase the value of $j$, the system approaches the classical limit and the signatures of chaos become clearer.

In order to further understand the signatures of chaos in the evolution of discord, we look at the time-averaged value of discord as we scan through different initial conditions. A contour plot of the time-averaged discord as a function of the initial spin coherent state clearly reproduces the regular and chaotic structures of the classical phase space (Fig. 3 ). Average discord is higher in the chaotic sea compared to the regular island. This is similar to signatures of chaos observed previously in time-averaged entanglement [11]. We also note that the average value of discord for initial spin coherent states in different parts of the chaotic sea reaches roughly the same

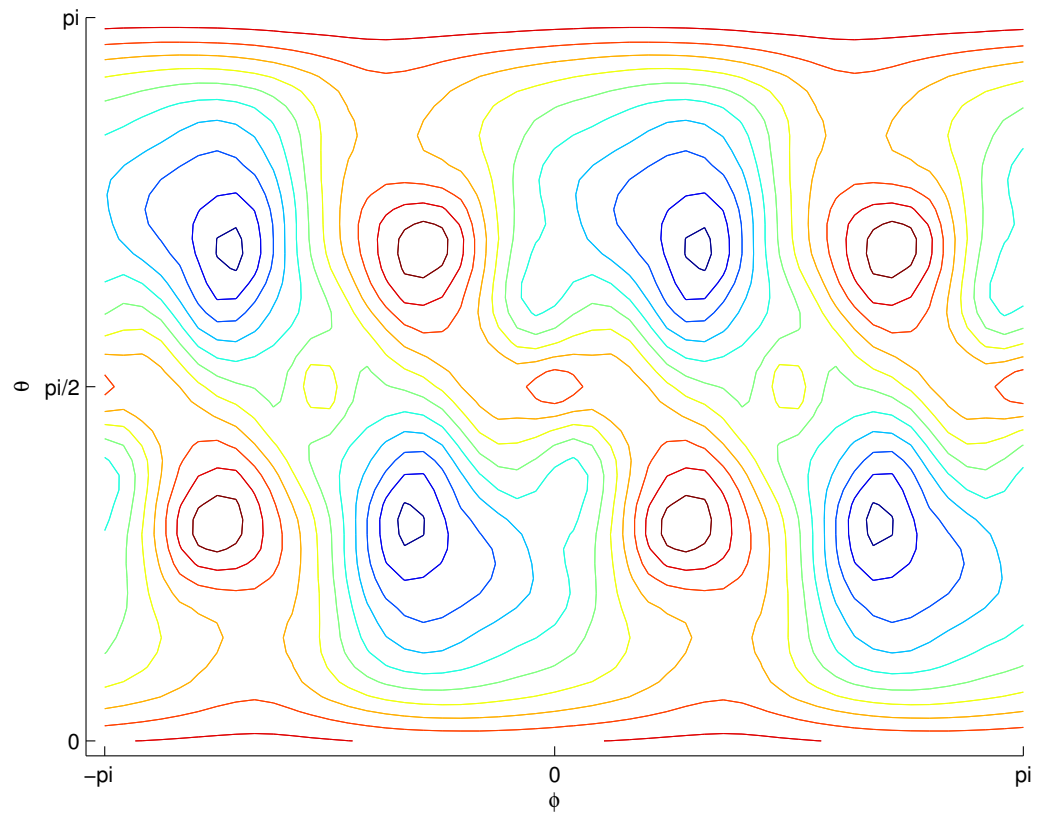

FIG. 3. Time-averaged discord as a function of the coordinates of the initial spin coherent state

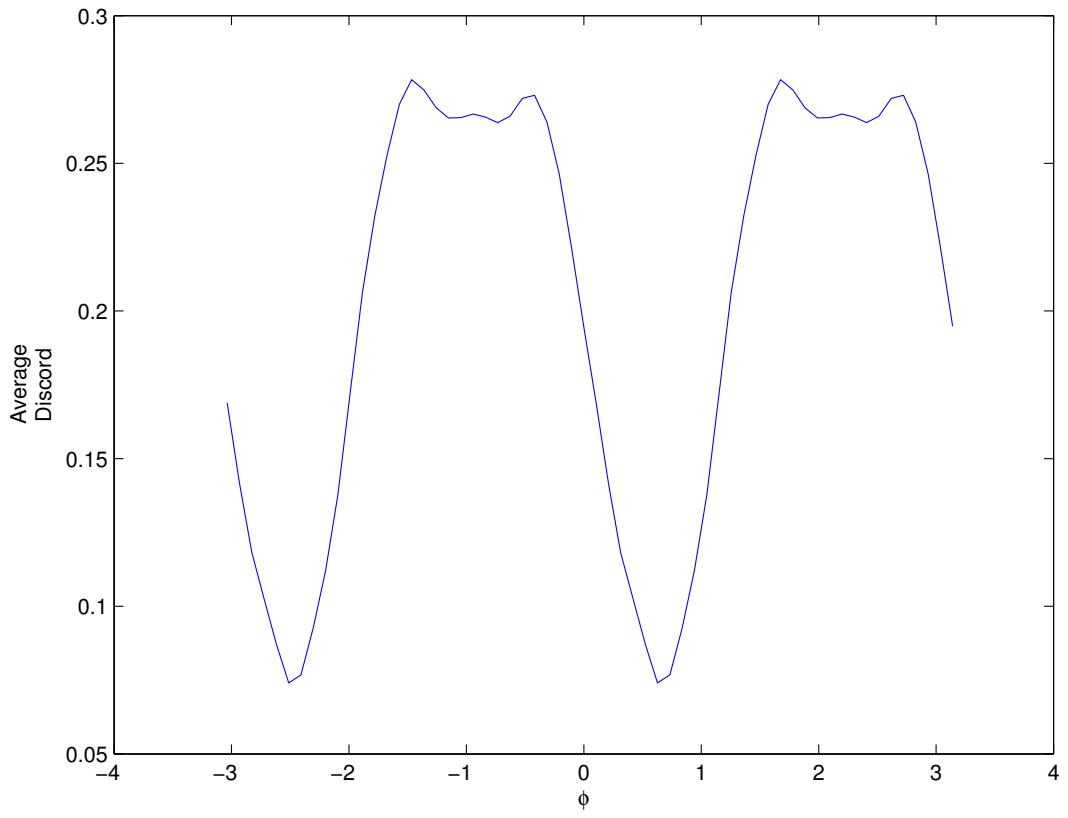

FIG. 4. Time-averaged 2-qubit discord as a function of the coordinates of the initial spin coherent state. We fix the polar angle $\theta$ and vary $\phi$ from -$\pi$ to $\pi$. The coherent state passes through two regular islands and the chaotic sea separating them. The long time average value of discord reaches a minimum near the fixed points and is almost flat in the chaotic sea. (here j=40)

value. To confirm this, we took a slice of the graph and plotted long time average discord for constant $\theta = 2.25$ and varying $\phi$ (Fig. 4). We see that the fixed point region has a significantly lower value of discord compared to the chaotic region. There is a sharp change as we cross from the regular islands into the chaotic sea. The average discord can thus be used to identify the edge of chaos.

To understand the nature of measures of correlations such as discord and entanglement and their relationship to each other, we next compare the discord dynamics with the entanglement dynamics. The two-qubit discord quantifies the correlation of these two qubits among themselves, while the two-qubit Von Nuemann entropy quantifies the entanglement of these two qubits with the rest of the (N-2) qubits. The two qubit concurrence quantifies entanglement of these two qubits with each other. Figures (5) and (6) show discord dynamics compared to Von Nuemann entropy and concurrence dynamics respectively. We find that discord dynamics mirror the entropy dynamics very well and behaves opposite to that of concurrence dynamics. When concurrence is high, discord is low and vice versa. Our calculations show that although both concurrence and discord are measures of quantum correlations, they are two separate quantities. Note that we can think of the remaining qubits as acting as an environment for the two qubits. Our results indicate that whereas concurrence undergoes sudden death, discord is more robust to an environment in the presence of chaos. Previous studies [27] have shown that discord is more robust than concurrence in Markovian environments, but the effect of chaos on this was not previously considered. Our results confirm the robustness of discord in a chaotic environment.

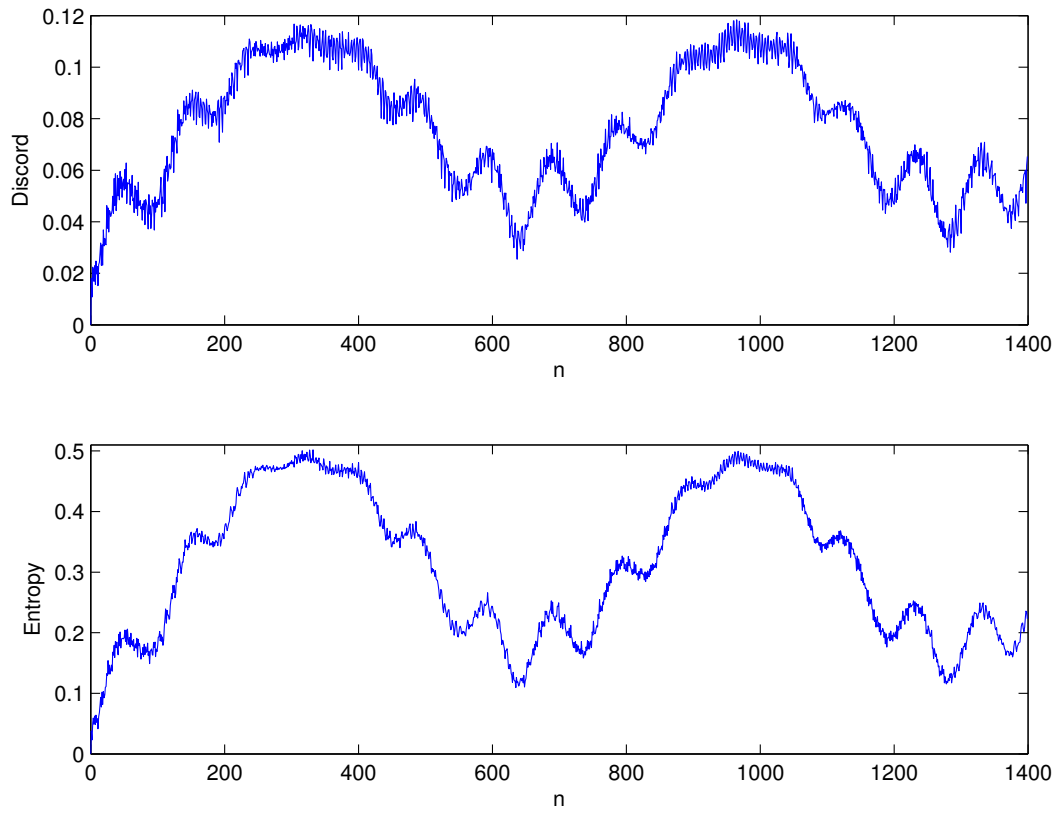

FIG. 5. Comparison of two-qubit discord and two-qubit entropy dynamics for $j = 4$. Discord dynamics perfectly mirrors entanglement between the two qubits and the remaining qubits

## DISCUSSION AND SUMMARY

The quantum kicked top is a simple but versatile system for studying various aspects of quantum chaos. A major advantage of this system is that it simulates a collection of $N = 2j$ qubits evolving in the symmetric subspace under exchange of qubits. This allows the possibility of studying different measures of quantum correlations such as entanglement and discord all in the same system. The finite dimension of the Hilbert space makes it possible to perform accurate calculations without errors introduced due to truncations issues. In this paper we have shown that the dynamics of two-qubit discord in a multiqubit system collectively evolving as a quantum kicked top shows signatures of chaos. The discord between any two qubits shows quasi-periodic modulations for initial states localized in regular regions. The periodic oscillation is lost when the initial state lies in the chaotic sea, and the discord rapidly rises to an almost constant value. The time averaged discord is higher when the initial conditions correspond to chaotic region of the classical phase space and the boundary between regular and choatic regions is sharply delineated by the change in the average discord.

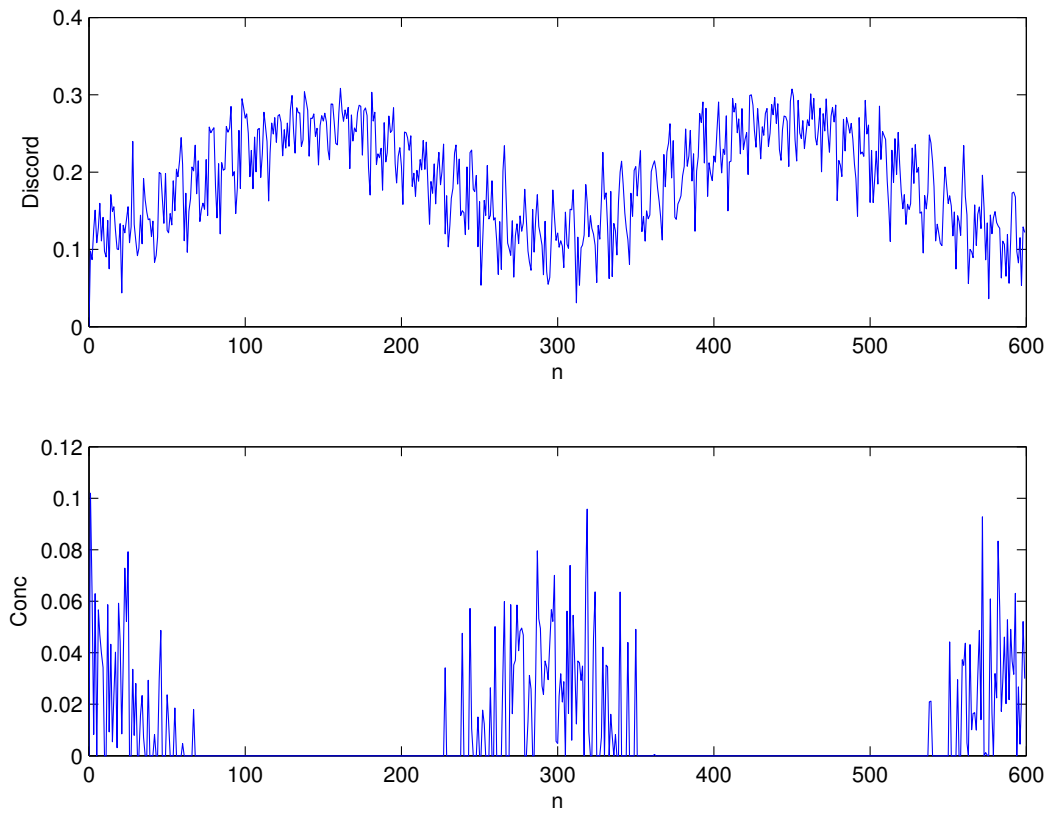

FIG. 6. Comparison of (a) two-qubit discord dynamics and (b) two-qubit concurrence dynamics for the $j = 4$ system. We find an inverse relationship between discord and concurrence.

Chaos occurs when the number of constraints or symmetries are fewer than the degrees of freedom. The same lack of symmetries at the quantum level means that the Hamiltonians cannot be described in block diagonal form. Instead, chaotic Hamiltonians have eigenstatistics that are well described by random matrices [2, 3]. Classical chaos can generate random probability distributions in phase space. Corresponding quantum dynamics can generate random states in Hilbert space [26]. When we focus our attention on the reduced subsystem of two qubits, this manifests as the generation of highly discordant states corresponding to the chaotic regions of the phase space. Quantum chaotic dynamics drives the system into arbitrary superposition of quantum states and this results in a higher average of value of discord in the chaotic part of the phase space as compared to the regular islands.

An interesting question that is relevant to quantum information processing applications is the comparison of various measures of quantum correlations- the two most important ones being entanglement and discord. Our calculations show that the two-qubit discord mirrors very closely the two-qubit Von Nuemann entropy, but behaves opposite to the two-qubit concurrence. Significantly, discord is more robust than concurrence in the presence of a chaotic environment. This raises interesting questions about the relationship of discord and concurrence, which we plan to explore in future work. Our results sheds new light on the behaviour of quantum correlations in chaotic systems, and since all parameters used are in an experimentally accessible regime, our work is relevant to future experiments exploring quantum chaos.

## ACKNOWLEDGEMENTS

We thank Ravinder R. Puri, Arul Lakshminarayan, Marco Piani, Kevin Resch, Alexei Kaltchenko and Animesh Datta for useful discussions. This research was supported by the Natural Sciences and Engineering Research Council of Canada, the Ontario Ministry of Research and Innovation, and Wilfrid Laurier University.

* vmadhok@gmail.com
† vibhugupta01@gmail.com